\RequirePackage{cmap}
\documentclass[reprint,superscriptaddress,aps,prl,showpacs,footinbib]{revtex4-1}

\usepackage{amsthm}
\usepackage{amsmath}
\usepackage{latexsym}
\usepackage{amsfonts}
\usepackage{amssymb}
\usepackage{color}
\usepackage{bbm,dsfont}
\usepackage{graphicx}
\usepackage{enumerate}
\usepackage{hyperref}
\usepackage{subfigure}
\usepackage{tikz}
\usepackage{placeins}
\usepackage[utf8]{inputenc}
\newcommand\opone{\leavevmode\hbox{\small1\kern-3.8pt\normalsize1}}

\DeclareUnicodeCharacter{00A0}{ }

\usepackage[normalem]{ulem}
\newcommand{\stkout}[1]{\ifmmode\text{\sout{\ensuremath{#1}}}\else\sout{#1}\fi}

\newtheorem{proposition?}{Proposition?}
\newtheorem{theorem}{Theorem}

\theoremstyle{definition}

\newcommand{\ket}[1]{|#1\rangle} 
\newcommand{\bra}[1]{\langle#1|} 
\newcommand{\kb}[2]{|#1\rangle\!\langle#2|} 
\newcommand{\tr}[2]{\text{tr}_{#2}\left\{#1\right\}}
\newcommand{\salg}{\mathcal{F}} 

\newcommand{\id}{\mathbbm{1}} 

\newcommand{\M}{\mathcal{M}}

\newcommand{\diff}{{\rm d}}
\newcommand{\norm}[1]{\vert\vert #1\vert\vert}
\newcommand{\abs}[1]{\vert #1 \vert}

\newcommand{\mean}[1]{{\mathcal M}\!\!\left[ #1 \right]}

\begin{document}

\title{{Continuous quantum measurement for general Gaussian 
  unravelings can exist}}

\author{Nina Megier}
\email{nina.megier@mi.infn.it}
\affiliation{Institut f{\"u}r Theoretische Physik, Technische Universit{\"a}t Dresden, 
D-01062,Dresden, Germany}
\affiliation{Dipartimento di Fisica “Aldo Pontremoli,” Università degli Studi di Milano, via Celoria 16, 20133 Milan, Italy}
\affiliation{Istituto Nazionale di Fisica Nucleare, Sezione di Milano, via Celoria 16, 20133 Milan, Italy}

\author{Walter~T.~Strunz}
\affiliation{Institut f{\"u}r Theoretische Physik, Technische Universit{\"a}t Dresden, 
D-01062,Dresden, Germany}

\author{Kimmo Luoma}
\email{kimmo.luoma@tu-dresden.de}
\affiliation{Institut f{\"u}r Theoretische Physik, Technische Universit{\"a}t Dresden, 
D-01062,Dresden, Germany}



\date{\today}


\begin{abstract}
  Quantum measurements and the associated state changes are 
  properly described in the language of instruments. We investigate the 
  properties of a time continuous family of instruments associated 
  with the recently introduced {family of} general Gaussian non-Markovian stochastic Schr\"odinger equations.
  In this Letter we find that when the covariance matrix for the Gaussian 
   noise satisfies {a particular $\delta$-function constraint,} 
   the measurement interpretation is possible
   for a class of models with self-adjoint coupling operator. This class contains,
   for example the spin-boson and quantum Brownian motion models with 
   colored bath correlation functions. Remarkably,
   due to quantum memory effects the reduced state,  in general, does not have 
   a closed form master equation while the unraveling has a time continuous 
   measurement interpretation.

  \end{abstract}

\maketitle

\paragraph{Introduction.---}{
{Open quantum system dynamics and quantum measurement 
 have a lot in common; both are described in terms 
of completely positive maps when the 
system and the environment or the system and measurement apparatus 
are initially prepared into a product state~ \cite{nielsenbook}. 
Since practically 
any quantum system is  coupled to an environment\cite{breuerbook}, 
detailed understanding of the 
intricate connections between the two physical processes is necessary
for both fundamental and applied research. 
For example, proper understanding of 
relaxation of a driven atom inside a leaky cavity  is provided by the theory of 
open quantum systems~\cite{Alicki:1105909} whereas the homodyning of the 
emitted light 
can be understood on the level of microscopical physical 
processes in terms of a time continuous measurement~\cite{carmichaelbook}.
Averaging over all possible measurement records in the latter case, provides the correct 
relaxation dynamics of the driven open system, thus reconciling the two
approaches.}

{More recently, the role of quantum measurements on the thermodynamical 
properties of open quantum systems have  been theoretically  
investigated~\cite{PhysRevLett.116.080403,Elouard2017,beyer2019steering}.
In current experiments, the individual conditional trajectories of continuously measured 
weakly coupled open quantum systems can be tracked~\cite{Murch2013,PhysRevLett.123.163601}.
With the eye on possible future experiments and quantum technologies, 
it is important to understand whether time continuous measurement interpretation 
is possible and how relevant conditional trajectories should be 
constructed   when the open system
is strongly coupled to its environment and possible memory effects are at 
play~\cite{diosi2014,PhysRevLett.120.150402}.}


An open quantum system can be studied in many 
different ways. One of the most widespread is the usage of {master equations 
for the} reduced
state, both in the time local \cite{breuerbook} and
time non-local form containing memory integrals \cite{nakajima, zwanzig}. 
In this article we focus on a different description, namely on 
unravelings of the reduced state evolution in terms of 
time continuous stochastic Schr\"odinger equations (SSEs). 
Applicability of such descriptions range from Markov evolutions \cite{gisin1992}, 
i.e. fulfilling the Gorini–Kossakowski–Sudarshan–Lindblad (GKSL) master equation \cite{doi:10.1063/1.522979,Lindblad1976}, to
highly non-Markovian dynamics {in terms of 
non-Markovian quantum state diffusion}~\cite{walter1996,walter1999}. The SSE 
formalism is well developed in the
Markov case, where a complete parametrization of {diffusive} SSEs
has been known already for some years \cite{wiseman2001,chia2011}. 
Similar progress in the non-Markovian regime has 
been made only recently
by the introduction of the general
Gaussian non-Markovian SSEs as well as the microscopically derived 
generalized Gaussian non-Markovian SSEs
~\cite{diosi2014,budini2015,ferialdi2016,PhysRevLett.120.150402,2018arXiv180403742G}.
  
{One} advantage of stochastic descriptions of the dynamics in 
the Markov regime lies in their physical interpretation. A 
single trajectory corresponds to an evolution that is conditioned
on time continuous monitoring of the environment of the 
open system~\cite{weber2014}. {For non-Markovian diffusive trajectories
driven by complex valued colored Gaussian noise, so far only a single-shot measurement
interpretation has been established
\cite{gambetta2003,lajos2008,wiseman2008,lajos2008er,walter2012,PhysRevA.61.062106}.
Interesting proposals for time continuous quantum measurements 
in presence of memory effects were given in \cite{lajos2008,lajos2008er}, 
where initially entangled observables were measured. However, it was pointed 
out subsequently in~\cite{wiseman2008} that such an approach 
cannot lead to pure state trajectories, thus strengthening the commonly held
viewpoint that in the presence of memory effects such a continuous measurement 
interpretation is not possible.

Remarkably, in~\cite{Barchielli_2010}{, using a complementary approach to ours,} 
the authors show that linear 
diffusive SSE driven by a real valued, {\it non-white} Ornstein-Uhlenbeck process leads to 
a random unitary type dynamics which does have a time continuous 
measurement interpretation. On the other hand, our approach, which will be elaborated on later, has a 
clear microscopic origin and the noise driving the process must only satisfy a
$\delta$-function constraint which we will give later in this Letter.}

The general Gaussian SSEs contain two types of correlation 
functions. Only the Hermitian (bath) correlation function 
$\alpha(t,s)$, occurring also in the standard SSEs, affects the reduced density operator evolution. 
The non-Hermitian correlation function $\eta(t,s)$ 
influences only the properties 
of the stochastic trajectories. 
The {new freedom}
in the description of the open quantum system dynamics introduced by
the correlation $\eta(t,s)$ has already turned out 
to be beneficial for tasks where optimization
over different pure state decompositions is needed, 
such as entanglement detection, entanglement 
bounds and entanglement
protection \cite{viviescas2010,guevara2014,PhysRevLett.120.150402}.
Consequently, it is a natural idea to examine whether the stochastic 
trajectories in this general description 
have a time continuous measurement interpretation beyond 
the white noise limit.

\paragraph{General Gaussian non-Markovian SSEs.---}{
The 
general Gaussian non-Markovian SSEs {investigated}
in~\cite{diosi2014,budini2015,PhysRevLett.120.150402} reads
\begin{equation}\label{eq:SSE_pre1}
\begin{split} \frac{\diff}{\diff t}\ket{\psi_t(z^*))} =
-iH_S\ket{\psi_t(z^*)} + L z^*_{t}\ket{\psi_t(z^*)}\\ 
- \int_{0}^t
\! \diff s \left[\alpha(t,s)L^\dagger+\eta(t,s)L\right]
O(t,s,z^*)\ket{\psi_t(z^*)},
\end{split}
\end{equation} where the {stochastic}  states $\ket{\psi_t(z^*)}$ are not
normalized and the initial state is {the same for all trajectories}:
$\ket{\psi_0(z^*)}=\ket{\psi_0}$. $H_S$ is a Hamiltonian of the open
system, $L$ is an operator describing the coupling of the open quantum
system to its environment. We have already assumed that the
assignment $\frac{\delta}{\delta
z^*_{s}}\ket{\psi_t(z^*)}=O(t,s,z^*)\ket{\psi_t(z^*)}$
is possible and $\overline{O}(t,z^*)=\int_0^t\diff s\, \alpha(t,s)O(t,s,z^*)$.
$O(t,s,z^*)$ can be calculated exactly for
many relevant systems, otherwise one has to turn to some approximation scheme
\cite{PhysRevA.60.91}.

The functions $\alpha(t,s)$ and $\eta(t,s)$ are, respectively,
the Hermitian and non-Hermitian correlation functions of the Gaussian complex
noise $z_t^*$ completely specified by its {mean  $\M[z_t^*]=0$ and 
correlations}
\begin{align}
\M[z_tz_s^*]=\alpha(t,s),\quad\M[z_t^*z_s^*]=\eta(t,s),
\end{align}
where the averages are taken with respect to the corresponding
Gaussian probability density. 

By construction, the
reduced density operator $\rho$ evolves according to a
completely positive and trace preserving map $\Phi_t$ obtained by 
averaging over the trajectories
obeying~\eqref{eq:SSE_pre1}; $\rho(t)= \M[\ket{\psi_t(z^*)}\bra{\psi_t(z^*)} ]=\Phi_t(\kb{\psi_0}{\psi_0})$
. Indeed, the family of stochastic pure states 
$\{\ket{\psi_t(z^*)}\}_{z^*}$ unravel the reduced evolution.
Rather surprisingly, the averaged reduced dynamics depends only on the
Hermitian correlation $\alpha(t,s)$, whereas the dependency on the
non-Hermitian correlation $\eta(t,s)$ occurs only in the trajectories
$\ket{\psi_t(z^*)}$.

The most general form for the Gaussian non-Markovian SSE is obtained
when the functions $\alpha(t,s)$ and $\eta(t,s)$ are only constrained
by a general positivity condition, which guarantees that they are
correlation functions for some complex Gaussian noise $z_t^*$
\cite{diosi2014,budini2015}.  

\paragraph{Measurement interpretation.---}{{
To investigate if the measurement interpretation of some stochastic process is possible, 
one introduces a notion of instrument and the measurement record~\cite{Holevobook,Barchiellibook}, 
see also Supplemental Material \footnote{See Supplemental Material at
\url{http://link.aps.org/supplemental/...} for more detailed
discussions , which includes
Refs. \cite{loubenets2001,Barndorff-Nielsen2002,Buchleitnerbook,
Ozawa1984,TN_libero_mab21108276,Pauwels1973,Picinbono,doob1953stochastic,novikov}
.}. 

Let $(\Omega_u,\salg_u)$ be a family of measurable spaces parametrized
by time $u\geq 0$.  Here  $\Omega_u$ is the set of all possible 
measurement records and $\salg_u$ is a family of increasing $\sigma$-algebras 
($\salg_{u'}\subset\salg_u,$ when $ u'<u$)
containing all possible events verifiable by a 
time continuous measurements up to time $u$.
Any measurement scheme is then 
described by a family of instruments 
{$\mathcal{Y}=\{Y_u(\cdot)[\cdot]:
\mathcal{F}_u \times
\mathcal{T}(\mathcal{H}_S)\rightarrow\mathcal{T}(\mathcal{H}_S); u\geq 0\}$},
where each $Y_u(F_u)$ is an instrument, that is, a  linear, trace non-increasing,
normalized  and completely positive map 
from  trace-class operators to trace class operators. 
Furthermore, the normalization is given by $\tr{Y_u(\Omega_u)X}{}=\tr{X}{}$,
where $X$ is an arbitrary trace class operator.
Causality sets an additional constraint on $\mathcal{Y}$:
Probability of verifying an event $F_s$ has to be invariant with 
respect to measuring up to some later time $t>s$ ($\mathcal{F}_s\subset \mathcal{F}_t$)  and discarding the gained
information.
Consequently, the  following 'compatibility demand' has to be
satisfied for all $s:$ $0< s<t$:
\begin{align}\label{compatibility} 
\forall F_s \in \mathcal{F}_s: &&
(Y_s)^{\dagger}(F_s)[\opone]=(Y_t)^{\dagger}(F_s \times
\Omega_s^t)[\opone],
\end{align} where $(Y_s)^{\dagger}(F_s)[\cdot]$ refers to the dual map
of $Y_s(F_s)[\cdot]$ and $F_s \times \Omega_s^t$ denotes all elements
of $\Omega_t$ which coincide with  $F_s$.

In this Letter, we choose that the complex
noise $z^*_t$ up to time $t$ is itself a measurement signal. In general, this is the case when the operator $O(t,s,z^*)$ occurring in the general Gaussian SSE \eqref{eq:SSE_pre1} has
at most linear dependence on $z^*_t$ \cite{walter2012}. However, we
will see later that this choice can always be made under exactly the same conditions  
when time continuous measurement interpretation exists. 
We denote by $G_t(z_t^*)$ the solution to Eq.~(\ref{eq:SSE_pre1}) 
with initial condition $G_0(z_0^*)=\id$. We can construct an instrument 
$Y_t(F_t)$ corresponding to general Gaussian SSE~(\ref{eq:SSE_pre1}) by setting
\begin{align}
  Y_t(F_t)(\kb{\psi_0}{\psi_0})=\int\limits_{F_t}G_t(z_t^*)\kb{\psi_0}{\psi_0}G_t^\dagger(z_t^*)\mu(\diff z_t),
\end{align}
which describes how the initial state $\ket{\psi_0}$ is mapped when 
measurement outcomes $z_t^*\in F_t$ are obtained. Here, $\mu(\diff z_t^*)$ is 
the Gaussian probability measure for the stochastic process $z_t^*$. 
Compatibility demand can be written in terms of the following 
two-times propagator 
\begin{align}
  A_s^t(z_t^*)=G_t(z_t^*)G_s^{-1}(z_t^*),
\end{align}
and with the help of Radon-Nikodym theorem \cite{billingsley,walter2012}
as follows  
\begin{align}\label{condition} \opone&=
\int\limits_{\Omega_s^t}A_s^{t\dagger}(z^*_t)A_{s}^t(z^*_t)
  \nu(\diff z^{*}_\tau|z^{*}_\sigma=\zeta_\sigma^*)\notag\\
  &\equiv\mean{A_s^{t\dagger}(z^*_t)A_{s}^t(z^*_t)\bigg|\zeta_\sigma^*},
  \,\, \nu_0^s-a.s.,
\end{align}
where the $\mean{\cdot|\zeta_\sigma^*}:=\mean{\cdot|\{z_\sigma^*| z_\sigma^*=\zeta_\sigma,\, \sigma\in (0,s]\}}$ 
is a shorthand notation for 
an expectation value conditioned on the history 
of a single noise realization $z^*_\sigma$ taking values 
$\zeta_\sigma$ from time $0$ till time $s$. 
Clearly, the condition (\ref{condition}) which guarantees that the probability of measuring a stochastic state is invariant with 
respect to measuring up to some later time $t>s$ and discarding the gained
information (causality) is equivalent to the the martingale condition
\begin{align}\label{eq:martingale}
  \mean{\norm{\psi_t(z_t^*)}^2\bigg|\zeta_\sigma}  =\norm{\psi_s(\zeta_s^*)}^2.
\end{align}
Accordingly, if the martingale condition \eqref{eq:martingale} is satisfied, the measurement interpretation is in principle possible.}

With the  preliminaries elaborated earlier, we can now 
state the main result of this Letter in the form of the the following theorem.
\begin{theorem}\label{thr:1}
If the correlations satisfy the following {$\delta$-function constraint}
\begin{align}\label{eq:corr_symm}
  \alpha(t,s)+\eta(t,s) = \kappa\delta(t-s),
\end{align}
and if the coupling operator $L$ is self-adjoint, a  
time continuous measurement interpretation is possible.
\end{theorem}

If condition (\ref{eq:corr_symm}) holds and $L=L^\dagger$, 
the SSE takes the simpler {time-convolutionless} form
\begin{align}\label{eq:SSEcond}
  \partial_t\ket{\psi_t(z^*)}=-iH_S\ket{\psi_t(z^*)}+z_t^*L-\frac{\kappa}{2}L^2\ket{\psi_t(z^*)},
\end{align}
since $O(t,t,z^*)=L$~\cite{PhysRevA.60.91}. {Clearly, $z_t^*$ can be taken 
as the measurement signal since how the operator  $O(t,s,z^*)$ would 
in general 
depend on $z_t^*$ has no influence on the stochastic dynamics when 
the $\delta$-function constraint (\ref{eq:corr_symm}) holds true.}
{Remarkably, under that constraint (\ref{eq:corr_symm}) 
the corresponding master equation for the reduced state is in general not closed 
}
\begin{align}\label{eq:gen_me}
  \partial_t\rho_t = &-i[H_S,\rho_t] +\Bigg(\mean{\overline{O}(t,z^*)\kb{\psi_t(z^*)}{\psi_t(z^*)}}L\notag\\
  &-L\mean{\overline{O}(t,z^*)\kb{\psi_t(z^*)}{\psi_t(z^*)}}+h.c.\Bigg),
\end{align}
{which  
can be seen as a signature of correlations between the open system and its environment 
which have a significant effect on the time scale of the open system evolution.} Significantly, the operator $O(t,s,z^*)$ occurs in the master Eq. \eqref{eq:gen_me}, as opposed to the case of SSE \eqref{eq:SSEcond}.

The master equation retains the Gorini-Kossakowski-Sudarshan-Lindblad (GKSL) form
when the time scale of the system-environment correlations is 
vanishingly small \cite{doi:10.1063/1.522979,Lindblad1976,PhysRevLett.120.150402} 
{and consequently the bath correlation is singular $\alpha(t,s)\propto\delta(t-s)$. 
In this case, the future of the stochastic process is independent of the past and conditional mean
values can be replaced by unconditional mean values~\cite{Note1}
. 
Accordingly, in the white noise limit the martingale condition (\ref{eq:martingale}) 
and the $\delta$-function constraint (\ref{eq:corr_symm}) are always satisfied \footnote{Here by 
white noise we mean that both $\eta(t,s)$ and $\alpha(t,s)$ are 
proportional to $\delta(t-s)$}.
Our $\delta$-function constraint  
given by Eq. \eqref{eq:corr_symm} allows, however, also for other forms of 
Hermitian correlation $\alpha(t,s)$, which, most notably, can also have finite correlation time.}

With this, we can proof the theorem~\ref{thr:1} as follows.
\begin{proof}
  {When the $\delta$-function constraint~ (\ref{eq:corr_symm}) holds, 
    the conditional mean value $\hat{\mu}_c(\tau)$ and the conditional 
    covariance $\Sigma_{c}(\tau,\tau')$ of $(z_\tau,z_\tau^*)^T$ with $\tau,\tau'\in(s,t]$ 
    satisfy
  \begin{align*}
  \hat{r}^\dagger \hat{\mu}_c(\tau)= 0, &&  \hat{r}^\dagger\Sigma_{c}(\tau,\tau')=\kappa\delta(\tau-\tau')\hat{r}^\dagger,
  \end{align*}
  where $\hat{r}=(r,r)$ and $r$ is an arbitrary operator acting 
  on the Hilbert of the system. If we additionally assume} that $L=L^\dagger$, the 
  conditional norm fulfills
  \begin{align*}
  \partial_\tau\mean{\norm{\psi_\tau(z^*)}^2\vert\zeta_\sigma} =0,
\end{align*}
  with initial condition 
  $\mean{\norm{\psi_s(z^*)}^2\vert \zeta_\sigma}=\norm{\psi_s(\zeta_s^*)}^2$. Thus 
  the norm squared is a martingale.
\end{proof}
We have presented the details of the proof in the 
Supplemental material~\cite{Note1}.

We want once again stress that the only conditions for the existence of a time continuous measurement interpretation of solutions of SSE \eqref{eq:SSE_pre1} are the self-adjointness of the coupling operator $L=L^\dagger$ and fulfilment of $\delta$-function constraint (\ref{eq:corr_symm}).
Consequently, the measurement interpretation is 
possible for important paradigms such as, among others, the spin-boson model, where 
$H_S=\frac{\omega}{2}\sigma_z$ and $L=g\sigma_x$ and quantum Brownian
motion, where $H_S=\frac{p^2}{2}+\frac{1}{2}\omega^2 q^2$ and 
$L=q$ even when the system-environment correlation function $\alpha(t,s)$ has a finite 
correlation time.

An important issue remains. Namely, the explicit construction of a non-trivial realisation of condition (\ref{eq:corr_symm}).}

\paragraph{Example: Ornstein-Uhlenbeck process.---}
A noise process satisfying condition~ (\ref{eq:corr_symm}) and 
having a  finite correlation time can be constructed using Ornstein-Uhlenbeck processes.
Suppose that the noise $z_t^*=x_t-i y_t$ driving the dynamics 
is decomposed into two independent real valued O-U processes $x_t,\,y_t$.
The noises $r_t$ satisfy $\dot r_t =-a_r r_t+\sqrt{D_r}\xi^{r}_t$, where 
$a_r > 0$ are the drift- and $D_r > 0$ the diffusion coefficients for   
$r=\{x,y\}$. 
$\xi^{x}_t$ and $\xi^{y}_t$ are uncorrelated standard real valued Gaussian white 
noises~\footnote{{$\mean{{\xi^\nu_t \xi^\sigma_s}}=\delta_{\nu\sigma}\delta(t-s)$}}.\\
Both real processes are of zero mean and the covariances 
are
\begin{align*}
  \langle x_t x_s\rangle = \frac{D_x}{2 a_x}e^{- a_x\abs{t-s}},\quad
  \langle y_t y_s\rangle = \frac{D_y}{2 a_y}e^{- a_y\abs{t-s}}.
\end{align*}
One can easily show that the Hermitian and non-Hermitian correlation functions read
\begin{align*}
  \alpha(t,s) = \langle x_tx_s\rangle +\langle y_ty_s\rangle, && \eta(t,s)=  \langle x_t x_s\rangle - \langle y_t y_s\rangle.                       
\end{align*}
When we take the limit $a_x\to\infty$ while $D_x/ a^2_x = \frac{\kappa}{2}$, the complex 
process $z_t$ satisfies the $\delta$-function constraint~(\ref{eq:corr_symm}). However, the Hermitian 
correlation function {$\alpha(t,s)$ has a finite correlation time} since it
 takes the form
\begin{align}
\alpha(\tau) = \frac{\kappa}{2}\delta(\tau) +\frac{D_y}{2 a_y}e^{- a_y\abs{\tau}},
\end{align}
which describes exponentially damped environment correlations.

\paragraph{Discussion.---}{
The question of the existence of a measurement
interpretation for non-Markovian quantum trajectories has 
a long and vivid history. After the introduction of the 
general Gaussian non-Markovian stochastic Schrödinger 
equations, new potential to explore this question has
opened up. In this paper we have investigated the possibility
to have a time continuous measurement interpretation for 
the non-Markovian trajectories {satisfying the relevant martingale condition (\ref{eq:martingale}).}

{We have shown that the stochastic trajectories 
of the  general Gaussian non-Markovian SSE can have a time
continuous measurement interpretation beyond the 
usual white noise limit. In fact, the stochastic trajectories of the 
general Gaussian SSE for models  with self-adjoint coupling operator and 
correlations satisfying the $\delta$-function constraint~(\ref{eq:corr_symm}) possess a
time continuous measurement interpretation.
Moreover, we also showed that the set of processes satisfying that
constraint is not empty.

{The framework we use is abstract and therefore
it has the advantage to describe the measurement process without
making any reference to a particular physical measurement setting. However,} an interesting question, which still has to be answered, is how such a
time continuous measurement can be implemented experimentally. 
{Accordingly,  we hope that our 
work inspires further investigations on suitable physical
interpretations of the stochastic trajectories described by
\eqref{eq:SSE_pre1} and the associated explicit experimental setup 
for time continuous measurement.}

\begin{acknowledgments}
N.M. acknowledges funding by the Alexander von Humboldt Foundation in form of a Feodor-Lynen Fellowship. 
\end{acknowledgments}

\bibliography{Draft-measurementNotes2}
\end{document}